\documentclass[journal=chematex,manuscript=article]{achemso}

\usepackage[version=3]{mhchem} 
\usepackage{gensymb}
\usepackage{chemformula} 
\usepackage[T1]{fontenc} 
\usepackage{textgreek} 
\usepackage{color,soul}
\usepackage[font=bf]{caption} 
\usepackage{lipsum}
\usepackage{multirow}
\usepackage{cuted}

\SectionNumbersOn 
\setkeys{acs}{keywords = false}
\setkeys{acs}{articletitle = true}

\newcommand{\llzo}{\ch{Li7La3Zr2O12}}

\author{Yuheng Li}
\affiliation[National University of Singapore]
{Department of Materials Science and Engineering, National University of Singapore, 9 Engineering Drive 1, 117575, Singapore}
\author{Asmee M. Prabhu}
\affiliation[]
{School of Chemical and Biomedical Engineering, Nanyang Technological University, Singapore 637459,
Singapore}
\author{Tej S. Choksi}
\affiliation[]
{School of Chemical and Biomedical Engineering, Nanyang Technological University, Singapore 637459,
Singapore}
\email{tej.choksi@ntu.edu.sg}

\author{Pieremanuele Canepa}
\affiliation[National University of Singapore]
{Department of Materials Science and Engineering, National University of Singapore, 9 Engineering Drive 1, 117575, Singapore}
\altaffiliation{Department of Chemical and Biomolecular Engineering, National University of Singapore, 4 Engineering Drive 4, 117585, Singapore}
\email{pcanepa@nus.edu.sg}

\title[]{\ch{H2O} and \ch{CO2} Surface Contamination of the Lithium-Stuffed Garnet}

\begin{document}


\begin{abstract}
Understanding the reactivity of ubiquitous molecules on complex oxides has broad impacts in energy applications and catalysis.
The garnet-type \llzo{} is a promising solid-state electrolyte for lithium(Li)-ion batteries, and it readily reacts with \ch{H2O} and \ch{CO2} when exposed to ambient air. Such reactions form a contamination layer on \llzo{} that is detrimental to the battery operations. The strong interactions of  \llzo{} with \ch{H2O} and \ch{CO2}, however, make \llzo{} a promising support to catalyze \ch{H2O} dissociation and \ch{CO2} adsorption.
Here, using first-principles calculations, we investigate the adsorption and reactions of \ch{H2O} and \ch{CO2} on a \llzo{} surface.
We show that \ch{H2O} reacts through the exchange of proton and \ch{Li^+} and produces metal hydroxide species. At high \ch{H2O} coverage, half of the \ch{H2O} molecules dissociate while the other half remain intact. \ch{CO2} reacts with the \llzo{} surface directly to produce carbonate species. We clarify that the individual reactions of \ch{H2O} and \ch{CO2} with \llzo{} are more thermodynamically favorable than the co-adsorption of \ch{H2O} and \ch{CO2}. Finally, we demonstrate that low temperature and high partial pressure promote the reactions of \ch{H2O} and \ch{CO2} with \llzo{}.
For energy storage application of \llzo{}, our study guides processing conditions to minimize surface contamination. From a catalysis point of view, our findings reveal the potential of using complex oxides, such as \llzo{} as a support for reactions requiring \ch{H2O} dissociation and strong \ch{CO2} adsorption.
\end{abstract}
\clearpage
\section{Introduction}

Energy storage and conversion devices relying on complex oxides are central to the decarbonization of our planet.\cite{dunnElectricalEnergyStorage2011,nykvistRapidlyFallingCosts2015,canoBatteriesFuelCells2018,warburtonTailoringInterfacesSolidState2021} 

On one hand, lithium(Li)-ion batteries based on oxide  chemistries,\cite{goodenoughChallengesRechargeableLi2010,dunnElectricalEnergyStorage2011} offer an appealing strategy to store green-energy produced by photo-voltaic panels, wind turbines, hydropower and/or hydroelectric means.  Li batteries suffer from safety issues, mostly due to the flammability of liquid electrolytes.\cite{dunnElectricalEnergyStorage2011} Solid-state batteries, replacing the flammable electrolyte with a solid ``ceramic'' ---solid electrolytes--- are safer alternatives.\cite{bachmanInorganicSolidStateElectrolytes2016,zhangNewHorizonsInorganic2018,janekSolidFutureBattery2016,famprikisFundamentalsInorganicSolidstate2019}
When suitably doped, the garnet-like quaternary oxide \llzo{} charts a solid electrolyte with unprecedented \ch{Li^+} conductivities ($\sim$10$^{-6}$--10$^{-3}$ S/cm).\cite{thangaduraiNovelFastLithium2003,muruganFastLithiumIon2007,buschmannStructureDynamicsFast2011,todaLowTemperatureCubic2013,galvenInstabilityLithiumGarnets2012,allenEffectSubstitutionTa2012,thangaduraiGarnettypeSolidstateFast2014,mukhopadhyayStructureStoichiometrySupervalent2015,chengInterrelationshipsGrainSize2015, sharafiImpactAirExposure2017,porzMechanismLithiumMetal2017,sharafiSurfaceChemistryMechanism2017,hanNegatingInterfacialImpedance2017,sharafiControllingCorrelatingEffect2017,canepaParticleMorphologyLithium2018} Importantly, \llzo{} is claimed to be stable against high-energy density Li-metal anodes, which is an appealing feature for high-energy density batteries enabling veichular transportation.\cite{porzMechanismLithiumMetal2017,sharafiControllingCorrelatingEffect2017,hanNegatingInterfacialImpedance2017,canepaParticleMorphologyLithium2018,kasemchainanCriticalStrippingCurrent2019} 

On the other hand, complex oxides catalyze the decomposition of abundant molecules, such as \ch{H2O} and \ch{CO2}. \cite{annamalaiInfluenceTightConfinement2018,lichtDFTInvestigationMechanism2017,brookesMolybdenumOxideFe2014,liHowControlSelectivity2019,trunschkeImpactBulkStructure2017} Water dissociation during water-gas shift is a vital elementary step that is promoted by  oxide supports,  such as \ch{Al2O3}, \ch{CeO_2}, \ch{CuO}, \ch{La2O3}, \ch{Mn2O4}, \ch{TiO_2}, \ch{Y2O3}, \ch{ZrO_2}, etc.\cite{liLowtemperatureWatergasShift2000,tanakaWaterGasShift2003,papavasiliouCombinedSteamReforming2007,rodriguezActivityCeOTiO2007,rodriguezWatergasShiftActivity2009,mcfarlandCatalysisDopedOxides2013,nelsonCarboxylIntermediateFormation2019}  Complex oxides are extensively studied as catalytic supports to adsorb \ch{CO2} from dilute gas streams and subsequently transform \ch{CO2} into valuable chemicals production.\cite{wangCarbonDioxideReforming1996,mcfarlandCatalysisDopedOxides2013,muroyamaCarbonDioxideMethanation2016,yangCombinedSteamCO22018,derkCatalyticDryReforming2014,wangHighlySelectiveStable2017,husCatalyticHydrogenationCarbon2019}

Here, using \llzo{} as an example, in the framework of \emph{ab initio} thermodynamics we investigate the reactivity  of this complex oxide with two ubiquitous molecules: \ch{H2O} and \ch{CO2}.
Previous work by \citeauthor{sharafiImpactAirExposure2017} using Raman and X-ray photo-emission spectroscopy (XPS)  revealed the existence of thick layers ($\sim$5-10~nm) of  \ch{Li2CO_3}  on the \llzo{} surfaces, upon exposure to air.\cite{sharafiSurfaceChemistryMechanism2017,sharafiControllingCorrelatingEffect2017} The same authors also detected \ch{LiOH} on the \llzo{} particles. The cartoon of Scheme~\ref{scheme1} shows the bulk \ch{CO2} and \ch{H2O} reactions with \llzo{} hypothesized by preliminary bulk reactions proposed by \citeauthor{sharafiImpactAirExposure2017}\cite{sharafiImpactAirExposure2017} using density functional theory (DFT).

\begin{scheme}[!ht]
\centering
\includegraphics[width=1.0\columnwidth,clip]{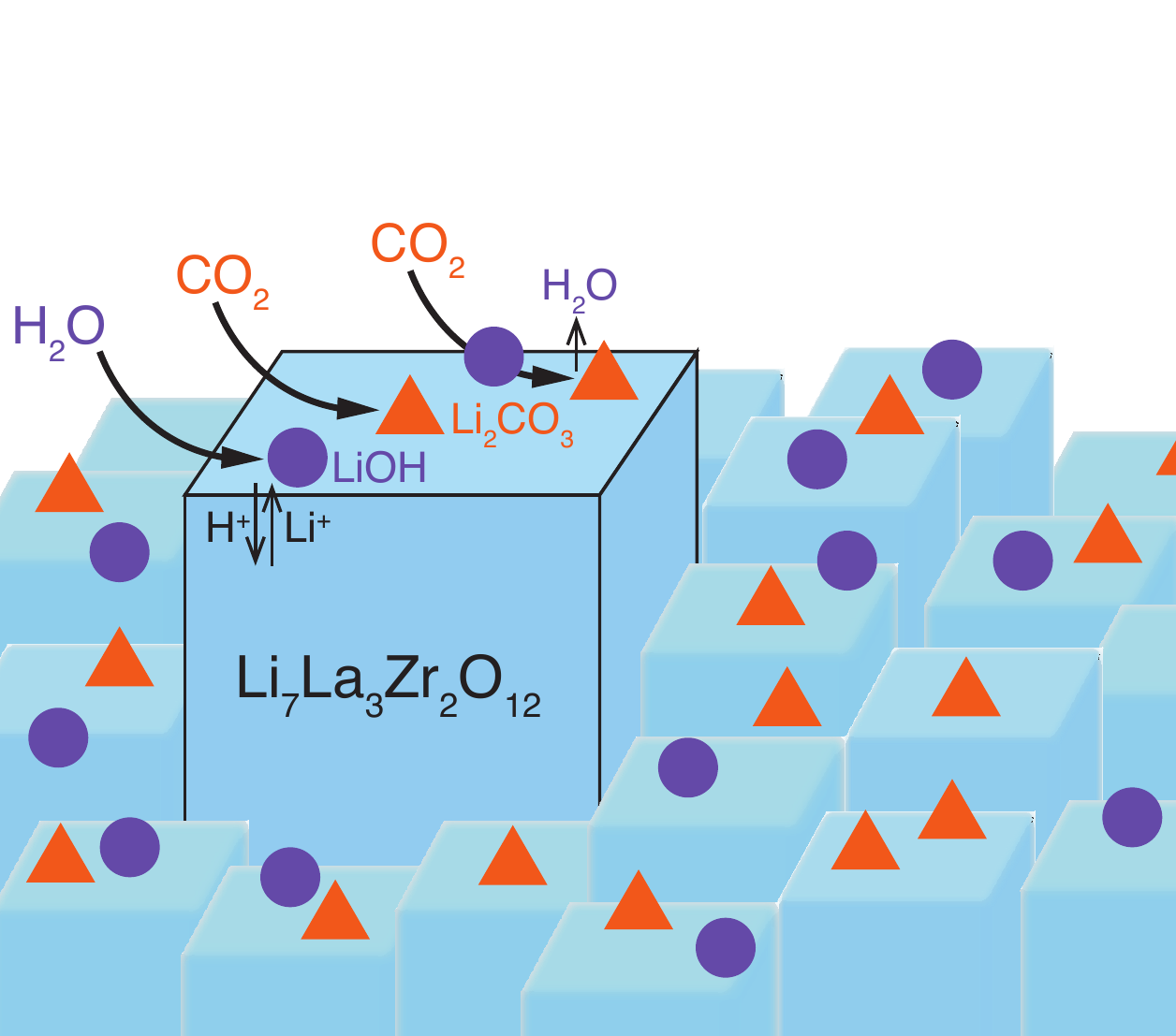}
\caption{Possible reaction pathways for \llzo{} particle contamination when exposed to \ch{H2O} and \ch{CO2}.}
\label{scheme1}
\end{scheme}

Nevertheless, the complexity of \llzo{} and other garnets in general, suggests that considering reactions of \ch{CO2} and \ch{H2O} beyond the bulk structure is necessary.\cite{sharafiImpactAirExposure2017} Specifically, if the structural heterogeneity of active sites on low index crystal planes contributes to activating closed shell species, such as \ch{CO2} and \ch{H2O}.  Here, we investigate this complex reactivity using DFT calculations of \llzo{} slabs interacting with \ch{H2O} and \ch{CO2}. The computed surface phase-diagrams reveal that at room temperature and ambient pressures \ch{H2O} adsorbs vigorously on the exposed \ch{Li} sites at the \llzo{} surfaces, nucleating \ch{LiOH} and protons. These protons are then exchanged with the highly-mobile \ch{Li^+}-ions and may "intercalate" in  the \llzo{}  bulk.\cite{galvenInstabilityLithiumGarnets2012} Likewise, our simulations demonstrate that \ch{CO2} adsorbs  on the \llzo{} surfaces and readily transforms into carbonate species. 

If \llzo{} is to be used in battery devices, our simulations provide guidelines to curb its degradation by \ch{CO2} and \ch{H2O}. In contrast, spontaneous \ch{H2O} dissociation on \llzo{} makes it potentially relevant for the water-gas shift reaction, \cite{zhaoImportanceMetaloxideInterfaces2017,nelsonHeterolyticHydrogenActivation2020} while the strong chemisorption of \ch{CO2} makes \llzo{} a promising dual functional material for \ch{CO2} capture and conversion. \cite{senHydroxideBasedIntegrated2020,omodolorDualFunctionMaterialsCO2020} Hence, in the context of catalysis, \llzo{} appears an excellent catalyst support to investigate reactions of \ch{CO2} activation and water splitting, respectively. These results are important to chart the reactivity of this complex quaternary oxide, which may be extended to other complex oxides.

\section{Computational Methods}
\label{sec:computational-methods}

The chemical reactions at the surfaces of \llzo{} were probed with Kohn-Sham
DFT simulations,\cite{kohnSelfConsistentEquationsIncluding1965} and approximated by the Perdew-Burke-Ernzerhof (PBE)  functional.\cite{perdewGeneralizedGradientApproximation1996} Van der Waals (vdW) interactions were treated with the D3 method with a Becke-Johnson (BJ) damping term.\cite{grimmeConsistentAccurateInitio2010,grimmeEffectDampingFunction2011,beckeExchangeholeDipoleMoment2007} The wavefunctions were expanded as plane-waves and  core electrons by the projector-augmented wave (PAW) method as in VASP.\cite{kresseEfficiencyAbinitioTotal1996,kresseEfficientIterativeSchemes1996,kresseInitioMolecularDynamics1993,kresseUltrasoftPseudopotentialsProjector1999} The PAW potentials and the recommended cutoffs energies were: C 08Apr2002 400 eV, H 15Jun2001 250 eV, La 06Sep2000 219 eV; Li\textunderscore sv 23Jan2001 272 eV; O 08Apr2002 400 eV; and Zr\textunderscore sv 07Sep2000 230 eV.
Slab optimizations were performed in two steps: \emph{i}) a preliminary PBE and $\Gamma$-point optimization, and \emph{ii}) optimization with PBE+D3-BJ with a 2$\times$2$\times$1 $k$-mesh and a cutoff energy of 440~eV. The total energy was converged to 1$\times$10$^{-5}$~eV/cell and forces  acting on atoms to within 0.01~eV/\AA{}.
We concluded with a final single-point energy calculations at a higher cut-off of 520~eV and integrated over a 4$\times$4$\times$1 $k$-point grid.


As in this study we probe the adsorption and reactivity of molecules at the surface of \llzo{}, it is crucial to treat explicitly the effect of vdW forces. vdW interactions were captured by the D3+BJ method.\cite{grimmeConsistentAccurateInitio2010,grimmeEffectDampingFunction2011,beckeExchangeholeDipoleMoment2007} Using these settings, the lattice constants ($a$~=~13.085~\AA{} and $c$~=~12.579~\AA{}) of the tetragonal ($Ia4_1/acd$) \llzo{} were used to rescale the (010) slab models from Ref.~\citenum{canepaParticleMorphologyLithium2018}. The (010) Li-terminated off-stochiometric \llzo{} model contains 248 atoms, and has a surface energy of $\sim$1.34~J/m$^2$.  The dispersion corrections (D3-BJ) increase the stability of the bulk structure, resulting in an increased surface energy by $\sim$0.50~J/m$^2$ compared to Ref.~\citenum{canepaParticleMorphologyLithium2018}. A well-converged vacuum size of 15~\AA{} was used to eliminate spurious slab-slab (and adsrobate-adsrobate) interactions along the direction orthogonal to the surface plane.

A bulk-like region of the slab where all atomic positions are fixed was introduced in the middle of the slab models. A 40\% bulk-like region was determined from convergence tests on slab total energies (see Figure S1 in Supporting Information) at which the total energy is converged to $\sim$7 meV and relatively modest computational costs.
Note, this procedure is commonly applied in surface science and catalysis to reduce significantly the computer time of structure optimization of large slab models. \cite{choksiPredictingAdsorptionProperties2019}
Benchmark tests on \llzo{} bulk lattice constants and surface energies were performed to compare the effects of  DFT functionals, i.e., PBE, PBE+D3 without damping, and PBE+D3 with BJ damping  (see Table S1 of SI).

The coordination numbers and Mulliken charges were derived from the crystal orbital Hamilton populations (COHP) available in the Lobster code.\cite{maintzLOBSTERToolExtract2016} For the COHP calculations, the energy range is set to --15$\sim$10 eV, and basis functions same as those in the PAW potentials are used for projections.

The adsorption energy ($\Delta E_\mathrm{ads}$) was calculated using Eq.~\ref{eq:1}: \cite{butlerDesigningInterfacesEnergy2019} 
  \begin{equation} 
  \label{eq:1}
  \Delta E_{\textnormal{ads}} = \frac{1}{2n} [E(\textnormal{\textnormal{slab}|2$n$\textnormal{H$_2$O$^{\textnormal{ads}}$}}) - E(\textnormal{slab}) - 2nE(\textnormal{H$_2$O$^{\textnormal{g}}$})],
 \end{equation}
where $n=1$ is the number of adsorbed molecules on each side of the slab model. In Eq.~\ref{eq:1} we approximated the Gibbs energy of each term by the DFT total energy (i.e., G~$\approx$~E), thus neglecting the zero point energy,  the $pV$ and entropy contributions. $E(\textnormal{\textnormal{slab}|2$n$\textnormal{H$_2$O$^{\textnormal{ads}}$}})$ is the total energy for the slab adsorbed with $n$\ch{H2O} (on each surface side), $E(\textnormal{slab})$ is total energy for the clean-surface slab, and $E(\textnormal{H$_2$O$^{\textnormal{g}}$})$ is total energy for \ch{H2O} in gas phase. From Eq.~\ref{eq:1}, a more negative $\Delta E_\mathrm{ads}$ value indicates a more favorable interaction of the adsorbate with the surface.

The energy changes computed as function of \ch{H2O} (and \ch{CO2}) coverage are calculated in terms of the Landau grand-potential energy, $\Omega$ of Eq.~\ref{eq:2_g}.
\begin{eqnarray} 
      \Omega (x/5 \ \mathrm{ML}) & = & \frac{1}{2A} \left [
      G\left(\mathrm{slab}|2x \mathrm{H_2O^{ads}} \right) - (12-2x) \mu^{T, P} (\mathrm{H_2O^{g}})
      \right], \label{eq:2_g}\\ 
     & \approx &  \frac{1}{2A} \left [ E\left(\mathrm{slab}|2x \mathrm{H_2O^{ads}}\right) - 2x T S_{vib}^T(\mathrm{H_2O^{ads}}) - (12-2x) \mu^{T, P} (\mathrm{H_2O^{g}}) \right], \label{eq:2}
\end{eqnarray}
where $x$ is the number of sites adsorbed on each surface, $A$ is the surface area, and $T$ and $P$ are temperature and partial pressure of \ch{H2O}, respectively. $G\left(\mathrm{slab}|2x \mathrm{H_2O^{ads}} \right)$ is the Gibbs free energy of the adsorbed slab and approximated as in Eq.~\ref{eq:2}, by $E(\mathrm{slab}|2x \mathrm{H_2O^{ads}})$ the DFT total energy of the slab adsorbing $x$\ch{H2O} molecules on each surface. $S_{vib}^T(\mathrm{H_2O^{ads}})$ is the vibrational entropy of the adsorbed \ch{H2O} (in either dissociated or intact form), calculated by fixing all atoms except the adsorbate. $\mu^{T, P} (\mathrm{H_2O^{g}})$ is the chemical potential of \ch{H2O} gas approximated by its DFT total energy in vacuum, corrected by the zero-point energy, and scaled to a given $T$ and $P$ condition using the Shomate equation.\cite{chaseNISTJANAFThermochemicalTables1998}  Then, the change of grand-potential energy $\Delta \Omega$ ($x$/5 $\mathrm{ML}$) at different water coverage is computed in Eq.~\ref{eq:3} with respect to the clean surface (0/5~ML). %
  \begin{equation} \label{eq:3}
      \Delta \Omega (x/5 \ \mathrm{ML}) = \Omega (x/5 \ \mathrm{ML}) - \Omega (0/5 \ \mathrm{ML}).
  \end{equation}
%
\section{Results}
\subsection{\llzo{} (010) Surface and Selection of Adsorption Sites}
\label{sec:surface}

Our analysis begins by selecting the energetically most favorable surface cut of \llzo{} from Ref.~\citenum{canepaParticleMorphologyLithium2018}. The off-stoichiometry Li-terminated (010) surface (and identical to the (100) and the (001) surfaces)  displays the lowest surface energy ($\sim$0.87$\pm$0.02~J/m$^2$) and remains stable even at high temperatures ($\sim$1000~K), targeted by typical thermal treatments of these oxides. The (010) surface corresponds to the dominant surface in the computed Wulff shapes of {\llzo{}}, and it is representative of the real {\llzo{}} particles, either in sintered powders or sheet composites. Here, off-stoichiometric surfaces refer to surfaces where the stoichiometry deviates in composition from the bulk. As described in the Computational Methods (Section~{\ref{sec:computational-methods}}), the (010) surface from Ref.~\citenum{canepaParticleMorphologyLithium2018} was rescaled and recalculated in this study with the incorporation of vdW interactions.

\begin{figure*}[!ht]
\centering
\includegraphics[width=1.0\textwidth,clip]{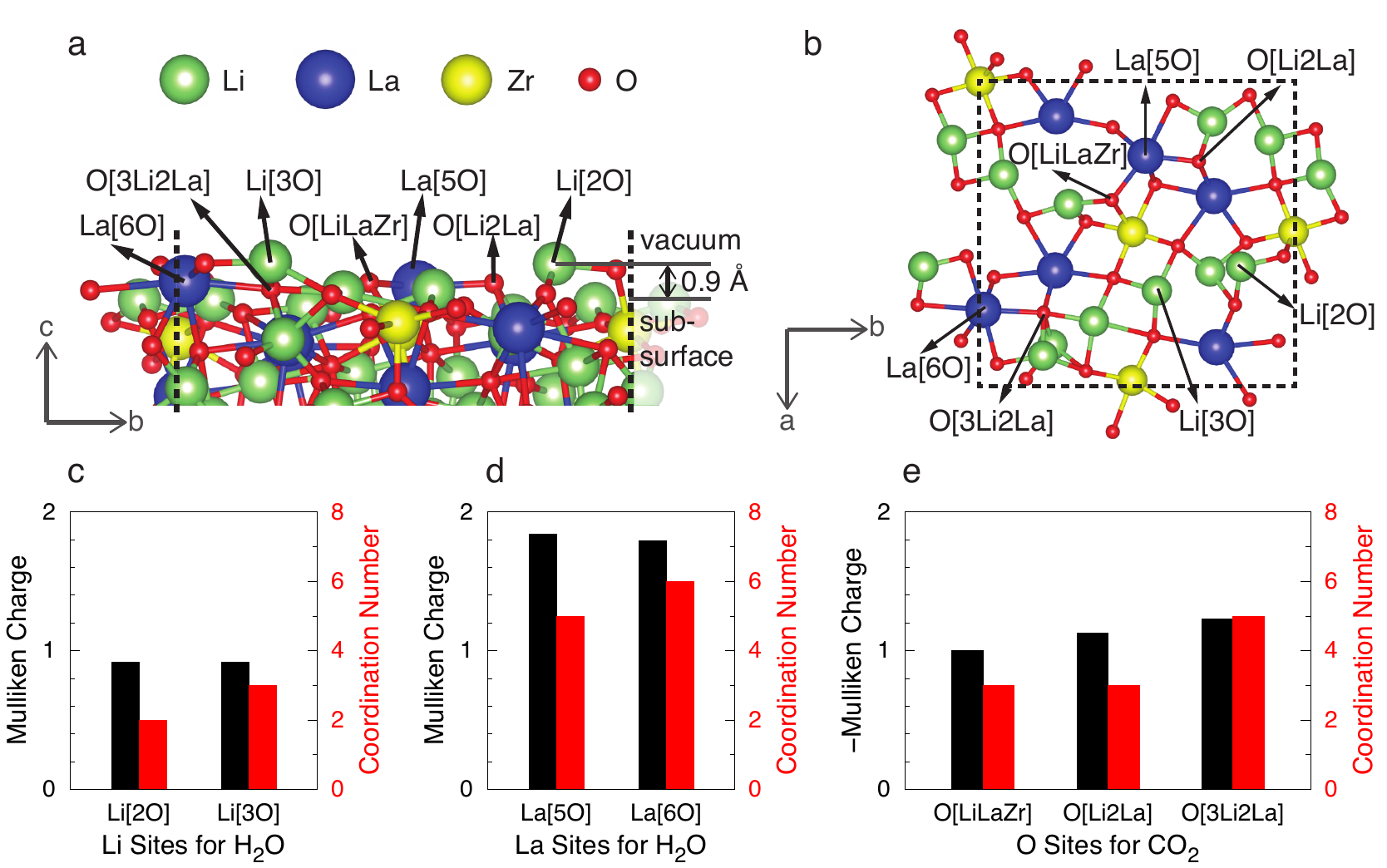}
\caption{Properties of the adsorption sites on a \llzo{} surface. (a) Side view and (b) top view of the off-stoichiometry Li-terminated (010) surface of \llzo{}. The boundary of the slab unit cell is marked with dashed lines. \ch{Li}, \ch{La}, \ch{Zr}, and \ch{O} species are shown in green, blue, yellow, and red, respectively. The surface adsorption sites are identified within $\sim$0.9~\AA{} vertical range (panel (a)). The double headed arrow marks the surface range, outside of which are the vacuum and subsurface. Sites are denoted by the element followed by ions immediately bonded to the site. The 7 unique sites are: 2 types of \ch{Li} (Li[3O] and Li[2O]), 2 types of \ch{La} (La[6O] and La[5O]),  3 types of \ch{O} (O[3Li2La], O[Li2La], and O[LiLaZr]), respectively. For a clearer view of  sites' local environments, we concealed atoms far away from the surface. Mulliken charges (black bars) and coordination numbers (red bars) for \ch{Li} sites in (c), \ch{La} sites in (d), and \ch{O} sites in (e).  \ch{Li} and \ch{La} sites show positive Mulliken charges, indicating their electro-positive character, and \emph{vice versa} for \ch{O} sites.}
\label{clean_surface}
\end{figure*}

When studying the adsorption of molecules on surfaces, a non-negligible challenge is selecting chemically sound adsorption configurations from the sheer number of imaginable adsorption conformations. This is especially true in complex oxides, such as \llzo{} which contains three chemically distinct cations with varying local coordination environments of oxygen atoms.  We used pymatgen  to ease the identification of unique adsorption sites, and subsequently construct initial structure of adsorbates bound to these sites.\cite{ongPythonMaterialsGenomics2013}  To prevent well-known convergence problems of slab calculations in the presence of fictitious electrical dipoles from polar adsorbates (\ch{H2O} here), we adsorbed molecules on both surface sides.\cite{taskerStabilityIonicCrystal1979} 

Tasked to clarify the mechanisms of \ch{H2O} and \ch{CO2} reactions with \llzo{}, it is first necessary to create appropriate adsorption models by understanding the  characteristics of different active sites on this structurally heterogeneous surface. These characteristics  include  the proximity of active sites to the \llzo{} surface, their local environment, their Mulliken charges, and their coordination numbers.

Figure~\ref{clean_surface}(a) and (b) show side and top views of the  off-stoichiometric Li-terminated (010) surface of \llzo.  By defining a range of search, along the non-periodic direction of the slab of $\sim$0.9~\AA{} from the most exposed atom at the surface, we identified 10 on-top surface adsorption sites. These sites show different proximity to surface; we denote each site according to its element followed by all species bonded to it. Thus, in Figure~\ref{clean_surface} seven unique sites out of 10 were identified:  two types of \ch{Li} sites Li[2O] and Li[3O]; two types of \ch{La} sites La[5O] and La[6O]; and three types of \ch{O} sites O[LiLaZr], O[Li2La], and O[3Li2La], respectively.

Nominally, positively charged ions, such as, \ch{Li^+} and \ch{La^{3+}} at the \llzo{} surface will attract negatively polarized parts of the adsorbing molecules and \emph{vice versa}. 

Figures~\ref{clean_surface}(c-e) show the computed Mulliken charges and coordination numbers for all seven types of adsorption sites. The Mulliken analysis is an intuitive (but not unique) way of repartitioning the electron charge density on each atom (and orbital) by projecting it onto individual orthonormalized atomic orbitals.\cite{szaboModernQuantumChemistry1996} The surface sites of each species show similar Mulliken charges, albeit distinct coordination numbers. As expected, \ch{Li} and \ch{La} metal sites show positive Mulliken charges. From an adsorption point of view, these sites will behave as Lewis acids in favor of accepting electrons from the oxygen  lone-pair of \ch{H2O}.  Indeed, the reactivity of these cations  follows the scale of absolute hardness, $\eta$: \ch{Li^+} ($\sim$35.1~eV) $>$ \ch{La^{3+}} ($\sim$15.4~eV) $>>$ \ch{Zr^{3+}} ($\sim$5.68 ~eV),\cite{parrAbsoluteHardnessCompanion1983,dronskowskiComputationalChemistrySolid2005} with \ch{Li^+} the most electrophilic cation of the three.  Since \ch{Zr^{3+}} is the least reactive metal of the three, we will not consider its reactivity in the remainder of this study. Oxygen sites reveals their basic character, and potentially driving the \ch{CO2} adsorption via the \ch{C} atom.  We use this knowledge to guide our understanding of the adsorption and dissociation of \ch{H2O} and \ch{CO2} on \llzo.

\subsection{\ch{H2O} Adsorption  and Hydroxide Formation on \llzo{}}
\label{sec:H2O}
Here, we provide the mechanistic insights of \ch{H2O} adsorption and reaction on the \llzo{} (010) surface to form hydroxide species.

Initially, we investigated the adsorption of a single \ch{H2O} molecule on the Li-terminated \llzo{} (010) surface.\ This low coverage analysis enables us to probe the surface characteristics of the surface.\ For every Li or La site, one \ch{H2O} was adsorbed on each side of the slab model, which eliminates any potential electrical dipoles caused by polar \ch{H_2O}.\cite{taskerStabilityIonicCrystal1979} The adsorption energies ($\Delta E_\mathrm{ads}$s) of Eq.~\ref{eq:1} were computed for all five Li and La  sites of Figure~\ref{clean_surface}.\cite{butlerDesigningInterfacesEnergy2019} 
  
Notably, the calculated $\Delta E_\mathrm{ads}$ indicate high propensity of a \ch{H2O} molecule toward \ch{Li^+} sites ($\Delta E_\mathrm{ads}$=--1.21~eV) compared to \ch{La^{3+}} sites (--0.85~eV), clearly following the scale of cation absolute hardness  by \citeauthor{parrAbsoluteHardnessCompanion1983}, with \ch{Li^+} ($\sim$35.1~eV) $>$ \ch{La^{3+}} ($\sim$15.4~eV).

\begin{figure}[!hb]
\centering
\includegraphics[width=1.0\columnwidth,clip]{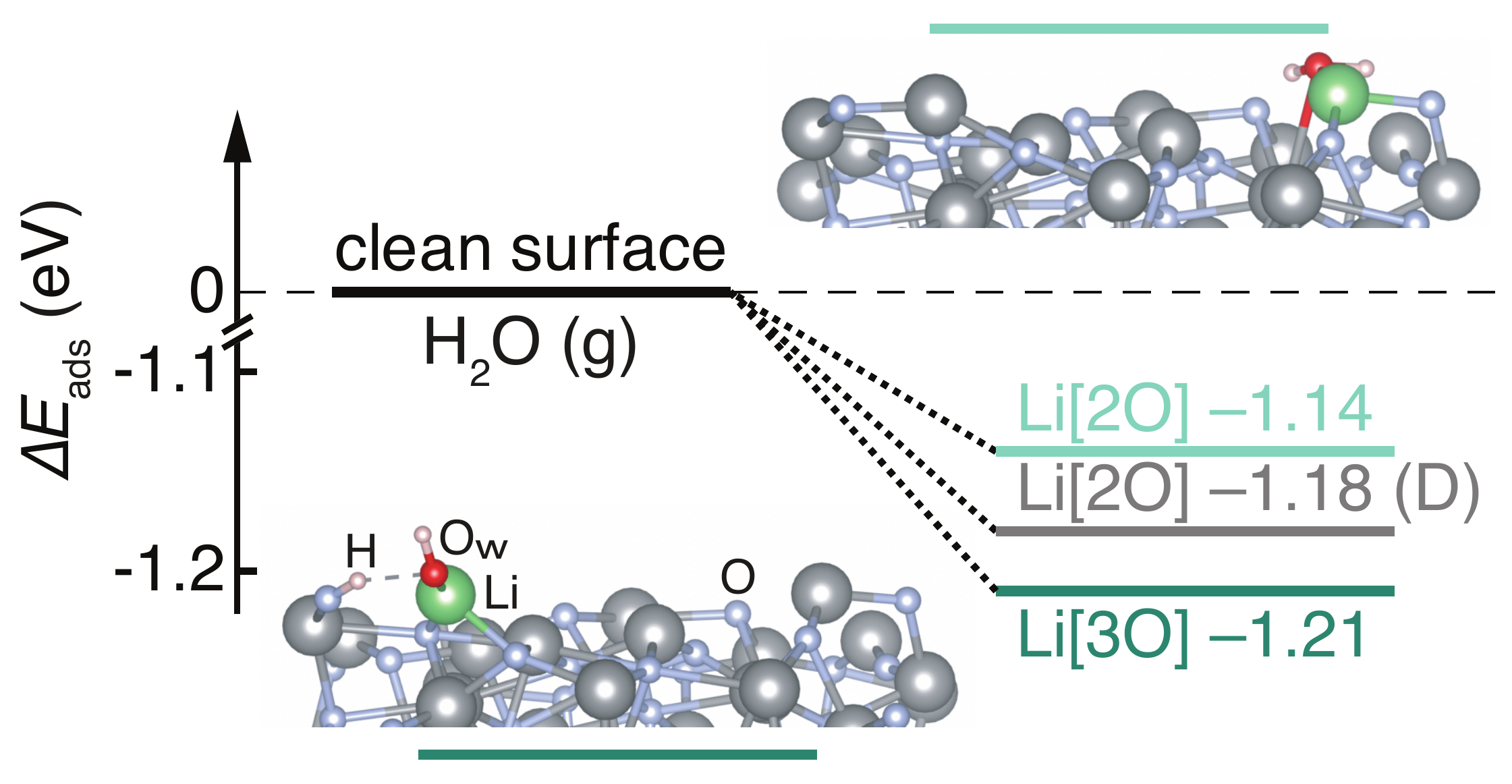}
\caption{
Adsorption energy for \ch{H2O} on the two most favorable sites Li[2O] (light green) and Li[3O] (dark green) on the Li-terminated \llzo{} (010) surface. The \ch{H2O} molecule on Li[2O] remains intact after adsorption (upper inset), while \ch{H2O} on Li[3O] dissociates to form LiOH (lower inset). The grey Li[2O] level marked with a ``(D)'' shows adsorption energy for a model where we forced \ch{H2O} dissociation on the Li[2O] site. Atomic species not directly involved in the adsorption are greyed out.
}
\label{single_H2O}
\end{figure}

Figure~\ref{single_H2O} shows the diagram of the computed adsorption energies and a magnification of the adsorption sites for the two most favorable adsorption cases, which always occur on exposed Li sites.
  
For the most favorable Li[3O] site (dark green line), \ch{H2O} dissociates accompanied by a significant $\Delta E_\mathrm{ads}$ ($\sim$--1.21~eV), and  reacts with the \ch{Li} site to form \ch{LiOH} (see lower inset of Figure~\ref{single_H2O}). Upon \ch{H2O} dissociation, a \ch{OH^-} group is formed between the proton released and an adjacent oxygen atom of the surface; the \ch{OH^-} group forms hydrogen bond with the O of the LiOH formed on the Li[3O] site. In the 2$^{nd}$ most favorable Li[2O] site (see upper inset of Figure~\ref{single_H2O}), the adsorbed \ch{H2O} remains intact, yielding still a non-negligible $\Delta E_\mathrm{ads}$ of $\sim$--1.14~eV, but comparatively less favorable than the Li[3O] site by $\sim$70~meV. Notably, we attempted the dissociation of \ch{H2O} on the Li[2O] site with a calculated $\Delta E_\mathrm{ads}$ of $\sim$--1.18~eV, indicating that the dissociation is thermodynamically  favorable by $\sim$--40~meV with respect to non-dissociative adsorbtion  at these sites.

It is important to unearth the causes behind the significant magnitudes of $\Delta E_\mathrm{ads}$s and dissociation behaviors of \ch{H2O} on different metal sites. We observed that if the coordinates of the surface sub-layers and the inner region of the surface (making $\sim$80\% of the slab) are held frozen to that of the bulk slab, the adsorbed \ch{H2O} will not dissociate on any site, regardless of their chemical activities. Clearly, with such a constraint in place, the adsorption energy does not benefit from the reorganization of the surface atoms, which promotes the dissociative adsorption of \ch{H2O}. Thus, the adsorption energy follows the trend of the site exposure to the surface, with the top (bottom) site showing the most negative adsorption energy. In contrast, by constraining only 40\% of the slab model, as in the rest of this study, the adsorbed \ch{H2O} on specific sites readily dissociates during the geometry optimization. Note that such preference for dissociation caused by rearrangements of atoms in the surface sub-layers is rarely seen in metals.

Unsurprisingly, \ch{H2O} dissociation lowers the adsorption energy. With a stabilizing contribution arising from \ch{H2O} dissociation, the adsorption energy no longer follows the trend of sites' proximity to the surface. From top to sub-layer sites in the surface, the adsorption energies are: --1.14~eV (Li[2O]), --1.21~eV (Li[3O]), --0.85~eV (La[5O]), --0.80~eV (La[6O]), and --0.65~eV (Li[3O]). Furthermore, our results clearly indicate that the dissociation behavior depends on local environment of the adsorption site. If a specific adsorption configuration displays oxygen atoms in the vicinity of the adsorption site, protons from adsorbed \ch{H2O} readily dissociate to bond with the surface oxygen species.
The dissociative adsorption of \ch{H2O} is well understood in previous studies of metal oxides including rutile-\ch{TiO2}, anatase-\ch{TiO2}, and rutile-\ch{RuO2}. \cite{wangProbingEquilibriumMolecular2017, calegariandradeFreeEnergyProton2020, nguyenDynamicsStabilityAdsorption2017}
In general, we find that adsorbed \ch{H2O} tends to dissociate when there are oxygen atoms within $\sim$1.9~\AA{} of the cation site (Table S2), unless there are metal ions that can better stabilize \ch{H2O} adsorption via bonding with O in \ch{H2O} (see inset for Li[2O] in Figure \ref{single_H2O}).
Therefore, values of $\Delta E _\mathrm{ads}$ in \llzo{} are clearly affected by \emph{i}) the site proximity to the surface and \emph{ii}) \ch{H2O} dissociation driven by the local environment. 

We are also interested in the bonding character of \ch{H2O} adsorption (or dissociation) on the \llzo{} surface since \ch{H2O}, a closed shell species, has significantly favorable $\Delta E_\mathrm{ads}$. First, we determined the dispersion contribution to the adsorption energy by comparing our predictions with (PBE-D3) and without  Van der Waals' contributions, hence just with GGA PBE.\cite{perdewGeneralizedGradientApproximation1996} The results show that there is a consistent dispersion contribution of only $\sim$15--20\% to the total DFT energies. This means that dispersion interactions are not a major source of bonding, and indicates a more dominant nature of chemisorption-driven reactions instead of physisorption. 

Next, we compared changes of Mulliken charges of  all  Li, La sites and O of \ch{H2O} before and after \ch{H2O} adsorption (Figure S3(a, b) in SI). Unlike on reducible oxides (with open-shell \emph{d} or \emph{f} transition metals), \cite{choksiPartialOxidationMethanol2016,metiuChemistryLewisAcid2012} there is no charge transfer between \ch{H2O} and the surface sites, which indicates the electrostatic nature of the interaction. These insights above agree with previous findings that intrinsic surface electrostatics in an insulating ionic system play a critical role in stabilizing the dangling lone pair of \ch{H2O}. \cite{kakekhaniNatureLonePairSurface2018}

\begin{table}[!ht]
 \caption{
 \ch{H2O} adsorption energy (in eV/molecule) at increasing number of molecules (ML). Site is the adsorption site.  $\mathbf{\delta \Delta E_{ads}}$ informs the change of $\mathbf{\Delta E_{ads}}$ upon the addition of a \ch{H2O} molecule. The number of dissociated \ch{H2O} molecules (\ch{H_2O}) are also indicated.  \ch{LiOH}  indicates whether the addition of a new \ch{H2O} molecule promotes \ch{LiOH}  formation. 
} 
\label{tab:loading}
  \begin{tabular}{clcccc}
  \hline
{\bf ML} & {\bf Site} & $\mathbf{\Delta E_{ads}}$ & $\mathbf{\delta \Delta E_{ads}}$ & {\bf \ch{H_2O}}  &  {\bf\ch{LiOH}}\\ 
\hline
1/5 & Li[3O] & --1.21 & --- &  1 & Yes \\
2/5 & Li[2O] &--1.19 & +0.02 & 1 & No\\ 
3/5 & La[5O]  &--1.19 & \,\,\, 0.00 & 2 & Yes\\  
4/5 & La[6O] &--1.10 &  +0.09& 2 & No\\ 
5/5 & Li[3O] &--1.09 & +0.01 & 3 & Yes\\ 
6/5 & Li[3O] &--1.07 &  +0.02 & 3 & No\\ 
 \hline
 \end{tabular}
\end{table}

We now investigate the adsorption of multiple \ch{H2O} molecules to simulate reactions at high water coverage on the \llzo{} surface. All five Li and La sites were first ordered according to their individual \ch{H2O} adsorption energy, from more negative to more positive. In particular, the order of adsorption energies is Li[3O] (--1.21~eV) $<$ Li[2O] (--1.14~eV) $\ll$ La[5O] (--0.85~eV) $<$ La[6O] (--0.80~eV) $<$ Li[3O]~(--0.65~eV). Following this order of stability, we adsorbed \ch{H2O} on available metals sites achieving a complete \ch{H2O} monolayer (ML). For each additional molecule, the structure was relaxed before the next \ch{H2O} was adsorbed. After all five sites were adsorbed, we identified an additional Li site, which emerged from the surface re-organization upon the increasing water coverage.  Eventually, six \ch{H2O} molecules were adsorbed on each side of the slab (12 in total). 

Table~\ref{tab:loading} reports the $\Delta E_\mathrm{ads}$ for \ch{H_2O} on the (010) \llzo{} surface as the exposed metal site are progressively saturated. We denote the \ch{H2O} coverage as the fraction of one monolayer ($x/5$~ML). $\Delta E_\mathrm{ads}$ for the same adsorptions but with all \ch{H_2O} forced to dissociate is reported in Table S3.

Table~\ref{tab:loading} shows that the addition of subsequent \ch{H2O} molecules to the \llzo{} (010) surface from 1/5 ML to 6/5 ML gradually increase $\Delta E_\mathrm{ads}$ to more positive values. This progressive increase is mainly caused by the more positive adsorption energy of each site, from --1.21 to --0.65 eV, dependent on their proximity to the surface and whether \ch{H2O} spontaneously dissociates. This increase is also likely due to additional \ch{H2O}--\ch{H2O} interactions as a result of crowding the \llzo{} surface, and the progressively neutralized and less reactive surface. Simultaneously, specific adsorption arrangements favor hydrogen bonding with the surface.
 

\begin{figure*}[!ht]
\centering
\includegraphics[width=1.0\textwidth,clip]{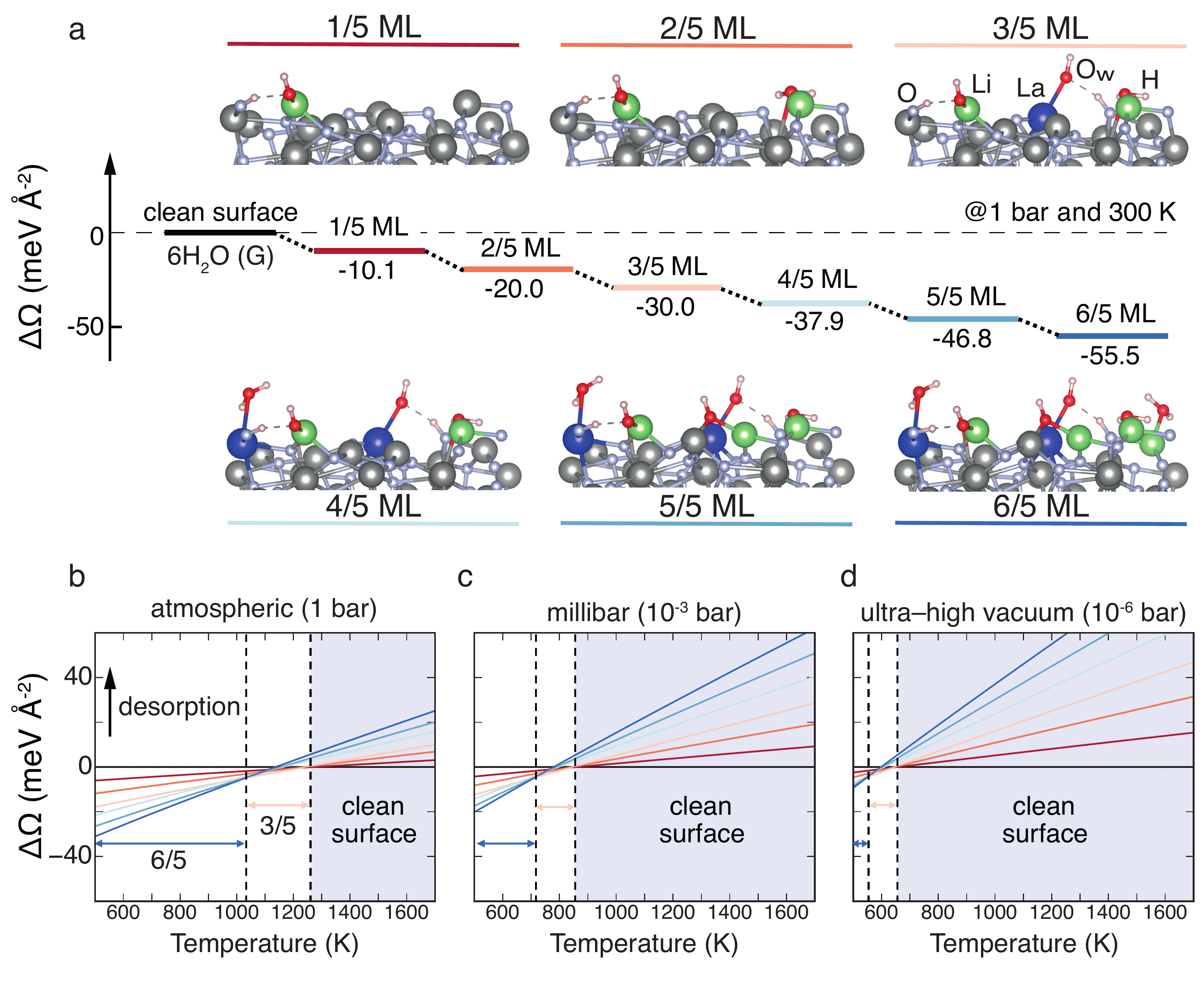}
\caption{Change of grand-potential energy $\mathbf{\Delta \Omega}$ relative to the clean \llzo{} surface and gas-phase H$_2$O at variable temperatures and pressures of technological relevance. In panel (a), the H$_2$O coverage is denoted by fraction of a monolayer (ML) of H$_2$O on the surface.   Full coverage is achieved at $\mathbf{6/5}$~ML when all the five original sites plus an additional site are adsorbed. Increasing coverage is marked by divergent colors from red to blue.  $\mathbf{\Delta \Omega}$ as a function of temperature for different H$_2$O coverage and partial pressure of 1 (b, atmospheric regime), 1$\times$10$^{-3}$ (c, mbar regime), and 1$\times$10$^{-6}$ bar (d, ultra-high vacuum regime). Double arrows show the most favorable coverage within temperature ranges. Atomic species not directly involved in the adsorption are greyed out.}
\label{multi_H2O}
\end{figure*}

Figure \ref{multi_H2O}(a) shows the energy diagram for all adsorption steps towards high \ch{H2O} coverage beyond one monolayer. $\Delta \Omega$ was calculated using Eq.~\ref{eq:2_g}--\ref{eq:3} for increasing \ch{H2O} coverage at the partial pressure of 1~bar and at 300~K. A negative $\Delta \Omega$ means that  \ch{H2O} adsorbs favorably on the \llzo{} surface. As more \ch{H2O} molecules adsorb, $\Delta \Omega$ continuously becomes more negative, indicating that the increasing coverage is favorable up to 6/5 ML. This trend also suggests that \ch{H2O}  can probably adsorb beyond 6/5 ML. We also found a decreasing trend for the absolute difference of $\Delta \Omega$ between  consecutive \ch{H2O} adsorptions. For example, the absolute difference between $\Delta \Omega$ of 6/5 ML and 5/5 ML is 8.7~meV~\AA$^{-2}$, less than 10.1~meV~\AA$^{-2}$ between 1/5 ML and 0/5 ML.
This decreasing difference is primarily attributed to the adsorption energy trends of individual sites, as less favored adsorption sites become occupied at higher coverages. 

As we only consider adsorption sites within $\sim$0.9~\AA{} from the surface top (bottom), more adsorption sites may be available. Electrostatic repulsion between adsorbates should have only secondary effects on the increasingly more positive values of $\Delta E_\mathrm{ads}$.  Furthermore,  a progressive addition of \ch{H2O} molecules contributes to gradually neutralizing the highly reactive \ch{Li^+} sites. Indeed, at  very high water coverage, beyond 6/5 ML, newly adsorbed \ch{H2O} molecules may not dissociate, even if dissociation is favored in a single \ch{H2O} adsorption on the same site. 

For coverage up to 6/5 ML, half of the adsorbed \ch{H2O} molecules dissociate and react with surface to form LiOH (see Table~\ref{tab:loading}). The same happens when all adsorbed \ch{H2O} molecules are forced to dissociate before relaxation (see Table S3). This ``half-dissociation'' trend was also observed for simple metal oxides like \ch{RuO2}, \cite{muDimerizationInducedDeprotonation2014, muDeprotonatedWaterDimers2015} but a different water dissociation fraction of 75\% was observed for the (101) surface in anatase-\ch{TiO2}. \cite{nadeemWaterDissociatesAqueous2018}
Note that the \ch{H2O} dissociation is spontaneous during geometry optimization, and thus it is not necessary to calculate their kinetic barriers explicitly (expected to be barrier-free).
In terms of hydrogen bonding for adsorbates, the dissociated proton always forms hydrogen bonds with the oxygen atom in the original \ch{H2O} molecule. All  \ch{H2O} adsorbates that remain intact form hydrogen bonds with adjacent surface O atoms.

Our analysis of the \ch{H2O} reactivity with the \llzo{} surfaces extends to conditions of technological relevance of this material. This analysis is relevant for the optimization of  synthesis and heat treatment conditions in \llzo{} and common oxides.  Figures \ref{multi_H2O}(b, c, and d) show $\Delta \Omega$ as a function of temperature at atmospheric (1~bar), millibar (10$^{-3}$~bar), and ultra-high vacuum (10$^{-6}$~bar) partial pressures, respectively.  The temperature range explored in Figures \ref{multi_H2O}(b, c, and d) span those of typical synthesis and sintering temperature ($\sim$1000 $^{\circ}$C) of \llzo{}.\cite{thangaduraiGarnettypeSolidstateFast2014}

At atmospheric conditions (Figure \ref{multi_H2O}(b)), $\Delta \Omega$ at different coverage is plotted in the same color scheme as in Figure \ref{multi_H2O}(a). A positive $\Delta \Omega$ indicates that adsorbed \ch{H2O} is in favor of desorption. We mark the most favorable coverage at different temperature ranges using double headed arrows. In general, as the temperature increases, the \ch{H2O} coverage  of the surfaces decreases until the surface of \llzo{} is water free. This situation is achieved for temperatures above $\sim$1260~K (Figure~\ref{multi_H2O} (b)).  At lower temperatures (up to $\sim$1030~K), the most favorable coverage is 6/5 ML \llzo, whereas the 3/5 ML coverage is the most favorable condition in the temperature range 1030--1260~K. Considering even higher coverage, one would expect steeper lines than for 6/5, resulting in water-rich surfaces of \llzo{}. However, such conditions cannot change the scenario for higher temperature ranges ($\geq$1260~K), exhibiting water-free \llzo{} surfaces. 
 
Under millibar (Figure \ref{multi_H2O}(c)) and ultra-high vacuum (Figure \ref{multi_H2O}(d)) \ch{H2O} partial pressures, one can find the same general trends for coverage vs.\ temperature as under atmospheric pressure (Figure \ref{multi_H2O}(a)).  However, the temperature ranges setting a water-free clean surface of \llzo{} increases as \ch{H2O} partial pressure decreases. For example, in situations of ultra-high vacuum (Figure \ref{multi_H2O}(d)), we predict an onset temperature of $\sim$650~K at which the \llzo{} surfaces will be water-free.  These findings indicate that low partial pressures are beneficial to minimize surface contamination with \ch{H2O} and \ch{LiOH}.


\subsection{\ch{CO2} Adsorption and Carbonate Formation on  \llzo{}}
\label{sec:CO2}
 We now focus on \ch{CO2} adsorption on \llzo{} and its subsequent transformation into carbonate \ch{CO_3^{2-}} species. We probed \ch{CO2} adsorption on five oxygen sites of the Li-terminated \llzo{} surface, as shown in Figure~\ref{clean_surface}. The adsorption is achieved by placing the carbon atom of \ch{CO2} close ($\sim$1.17~\AA) to the O site, to induce the formation of a \ch{C-O} bond. The most favorable site is a O[LiLaZr], and its $\Delta E_\mathrm{ads}$ is strikingly high $\sim$--2.21~eV ---a clear indication of \ch{CO2} chemisorption. The other sites show comparable ($\sim$--1.65--1.31~eV) but more positive adsorption energies. The least favorable site is O[3Li2La], still with a sizeable $\Delta E_\mathrm{ads}$ of $\sim$--1.31 eV.

\begin{figure*}[!ht]
\centering
\includegraphics[width=1.0\textwidth]{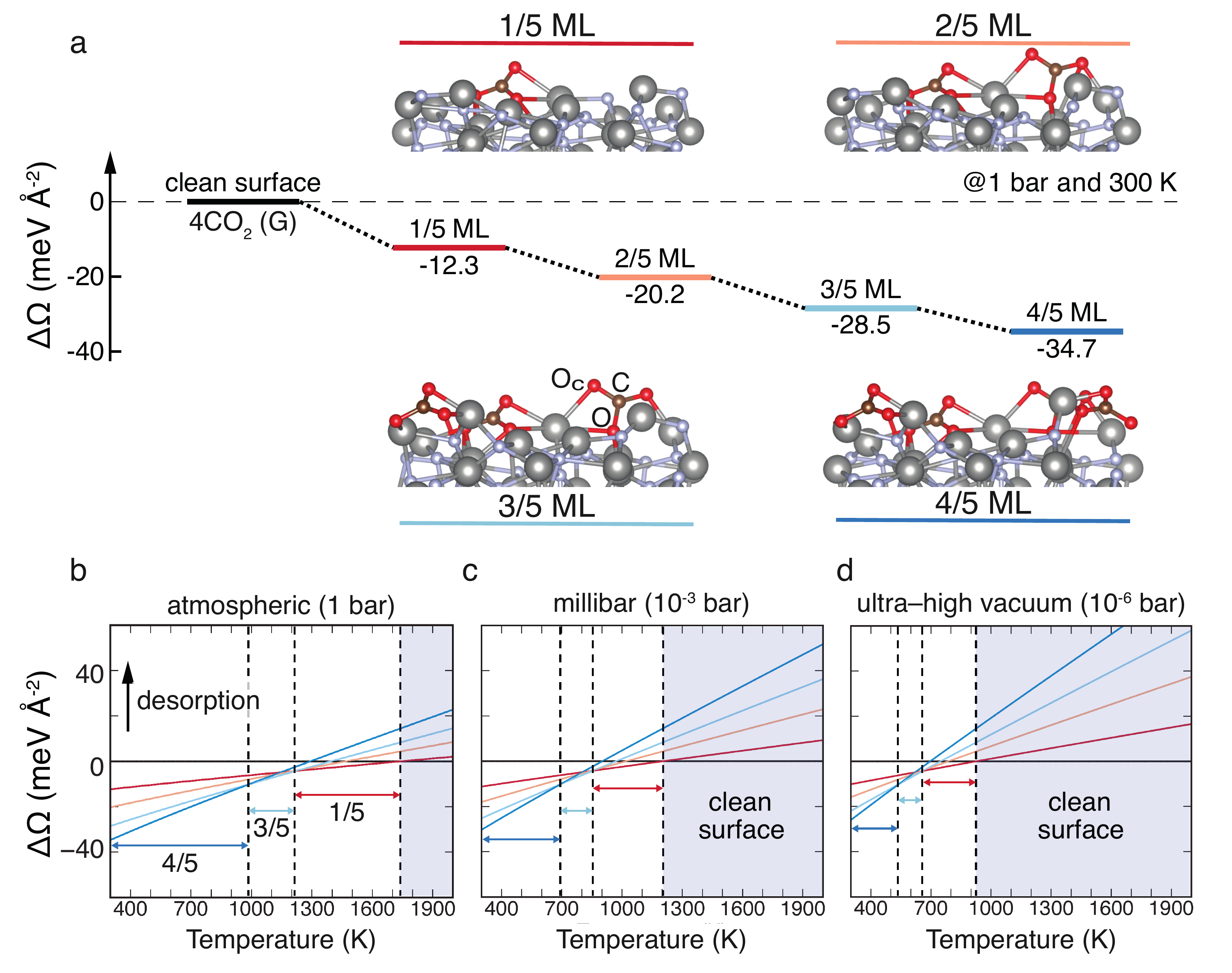}
\caption{
Change of grand potential $\mathbf{\Delta \Omega}$ relative to clean surface and gas-phase \ch{CO2} at variable temperatures and pressures of technological relevance. In panel (a), the \ch{CO2} coverage is denoted by a fraction of \ch{CO2} monolayer (ML) of the surface. $\mathbf{\Delta \Omega}$ (in meV/\AA$^{2}$) as a function of temperature for different \ch{CO2} coverage and partial pressure of 1 (b, atmospheric regime), 1$\times$10$^{-3}$ (c, millibar regime), and 1$\times$10$^{-6}$ bar (d, ultra-high vacuum regime), respectively. Double arrows denote the most favorable coverage within each temperature range. The structures and $\Delta \Omega$ for different coverage are marked using the same color scheme of four divergent colors. Atomic species not directly involved in the adsorption are greyed out.
}
\label{multiple_CO2}
\end{figure*}

Regardless of the binding oxygen site,  the adsorbed \ch{CO2} reorganize into carbonate \ch{CO_3 ^{2-}} species after structure optimization. In detail, two oxygen atoms in \ch{CO2} form bonds with metal ions available in the vicinity, with the carbonate species organizing in a polydentate conformation  (Figure S1).\cite{tsujiDynamicBehaviorCarbonate2003} For all carbonates formed on the five sites, there are 13 bonds formed between surface metal ions and O of \ch{CO2}, including eight Li-O bonds and five \ch{La-O} bonds. The exceeding number of \ch{Li-O} bonds show a clear preference for \ch{Li^+} over \ch{La^{3+}} ions. In particular, the surface carbonates show an average \ch{C-O} bond length of $\sim$1.31~\AA, comparable to $\sim$1.28~\AA{} in bulk \ch{Li2CO3}. \cite{idemotoCrystalStructureLixK11998}

The adsorption of \ch{CO2} consists of two types of bonds: one between the carbon atom of \ch{CO2} and a surface oxygen site,  and   the other between the oxygen atoms of \ch{CO2} and metal ions of the surface. We observe that the incipient bond between O and metal-ion on \llzo{} has major contributions to the values of $\Delta E_\mathrm{ads}$. On different adsorption sites, the strength for the O and metal ion bonding depends on which metal ions are proximal to the site.

We also determined the bonding nature for \ch{CO2} chemisorbed on \llzo{}. Indeed, the dispersion contributions are calculated to be only $\sim$15-20\%, similar to the cases of \ch{H2O} adsorption discussed earlier. Charge transfer is then analyzed through changes of Mulliken charges upon adsorption. Figure S3 shows that the average changes are $\sim$0.30 for the surface O sites and $\sim$--0.26 for C in \ch{CO2}. Besides, the average change upon adsorption is $\sim$--0.22 for O in \ch{CO2}, $\sim$--0.01 for surface Li ions bonded to O, and $\sim$0.12 for La ions. These changes are slightly larger than those for \ch{H2O} adsorption ($\sim$--0.05 for Li sites, $\sim$0.07 for La sites, and $\sim$0.03 for O in \ch{H2O}, see Figure S3 in SI). 
Taken together, this analysis confirms that the structural heterogeneity of surface sites on \llzo{} (010) enables strong adsorption of  \ch{H2O} and \ch{CO2}, which are otherwise known to physisorb on metals and ionic oxides. While such robust adsorption events are detrimental for the application of \llzo{} in energy storage, they indicate that \llzo{} is a potentially interesting catalyst support for \ch{H2O} activation and adsorbing \ch{CO2} from dilute streams.

We then studied \ch{CO2} adsorption as a function of  \ch{CO2} coverage. \ch{CO2} was sequentially adsorbed on all five O sites available on the \llzo{} surface. The sequential adsorption following the order of their individual adsorption energies, from more negative to more positive (O[LiLaZr] (--2.21~eV) $\ll$ O[Li2La] (--1.65~eV) $<$ O[LiLaZr] (--1.58~eV) $<$ O[LiLaZr] (--1.34~eV) $<$  O[3Li2La] (--1.31~eV)). In the process to adsorb the last \ch{CO2} onto the \llzo{} surface,  the O[3Li2La] site was found unfavorable,  as adsorbed \ch{CO2} on this site was too close ($\sim$1.20~\AA{}) to a \ch{La^{3+}} on the reconstructed surface. Eventually, there are only four out of five oxygen sites available for \ch{CO2} adsorption, which set the highest \ch{CO2} coverage to 4/5 ML. Besides the original five sites, less exposed O surface sites could increase \ch{CO2} coverage beyond 4/5 ML.

Table~\ref{tab:loadingCO2} shows the computed $\Delta E_\mathrm{ads}$ for increasing \ch{CO_2} loading up to the condition of 4/5 ML.

\begin{table}[!th]
  \caption{
  \ch{CO2} adsorption energy (in eV/molecule) with increasing coverage of molecules (ML). Site is the adsorption site. $\mathbf{\delta \Delta E_{ads}}$ informs the change of $\mathbf{\Delta E_{ads}}$ upon the addition of a \ch{CO2} molecule. \ch{CO_3^{2-}}  indicates whether the addition of a new \ch{CO_2} molecule promotes carbonate formation.  
} 
\label{tab:loadingCO2}
{
  \begin{tabular}{clccc}
  \hline
{\bf ML} & {\bf Site} &  $\mathbf{\Delta E_{ads}}$ &  $\mathbf{\delta \Delta E_{ads}}$ &   {\bf\ch{CO_3^{2-}}}\\ 
\hline
1/5 & O[LiLaZr]  &--2.21 & --- & Yes \\
2/5 & O[Li2La]  &--1.85 & +0.36 & Yes \\ 
3/5 & O[LiLaZr] &--1.75 & +0.10 & Yes \\  
4/5 & O[LiLaZr] &--1.62 & +0.13 & Yes \\ 
 \hline
 \end{tabular}
 }
\end{table}

Figure \ref{multiple_CO2}(a) shows structure schematics and $\Delta \Omega$ diagram for each adsorption step toward the 4/5 ML coverage. $\Delta \Omega$ is defined  as in Eq.~\ref{eq:2}-\ref{eq:3}, and here it is calculated at 300~K and a \ch{CO2} partial pressure of 1~bar. 

In regimes of high-coverage, adsorption of  multiple \ch{CO2} interact with the surface in the same way as singly  adsorbed  \ch{CO2}  molecules, transforming into \ch{CO_3^{2-}} units. At the highest calculated coverage of 4/5 ML, all adsorbed \ch{CO2} react with the surface and form carbonates. As the coverage increases, $\Delta \Omega$ continuously decreases, indicating that the reaction between the surface and an increasing amount of \ch{CO2} (up to 4/5 ML) is thermodynamically favorable. However, from Figure~\ref{multiple_CO2}, one observes that absolute difference of $\Delta \Omega$ between consecutive steps decreases progressively, similar to  trends of  Figure \ref{multi_H2O}(a).  Figure~\ref{multiple_CO2} and Table~\ref{tab:loadingCO2} show that each oxygen species at the \llzo{} surface become increasingly less reactive as the \ch{CO_2} coverage increases, which contributes to a progressive decrease of the $\Delta E_\mathrm{ads}$  and $\Delta \Omega$.

Figures \ref{multiple_CO2}(b-d) show $\Delta \Omega$ at technologically relevant temperature and \ch{CO2} partial pressures. As the temperature increases, the adsorption and reaction of \ch{CO2} become less favorable, and the preferred coverage decreases from 4/5 ML to conditions of \ch{CO2}-free surfaces. Similarly to the \ch{H2O} case, at low temperature even higher \ch{CO2} coverage than 4/5 ML may become accessible. However, this does not change significantly the situation of low \ch{CO2}-free surfaces at high temperature, as shown in Figure~\ref{multiple_CO2}(b-d). As the \ch{CO2} partial pressure decreases from atmospheric to ultra-high vacuum, the formation of carbonates species at the \llzo{} surfaces appear less favorable, with the stability ranges of temperature  expand accordingly. This indicates that low \ch{CO2} partial pressures are beneficial to minimize the carbonate contamination of  \llzo{} surfaces.

\subsection{\ch{CO2} and \ch{H2O } co-adsorption  on \llzo}
\label{sec:pathways}

\begin{figure}[!ht]
\centering
\includegraphics[width=1.0\columnwidth,clip]{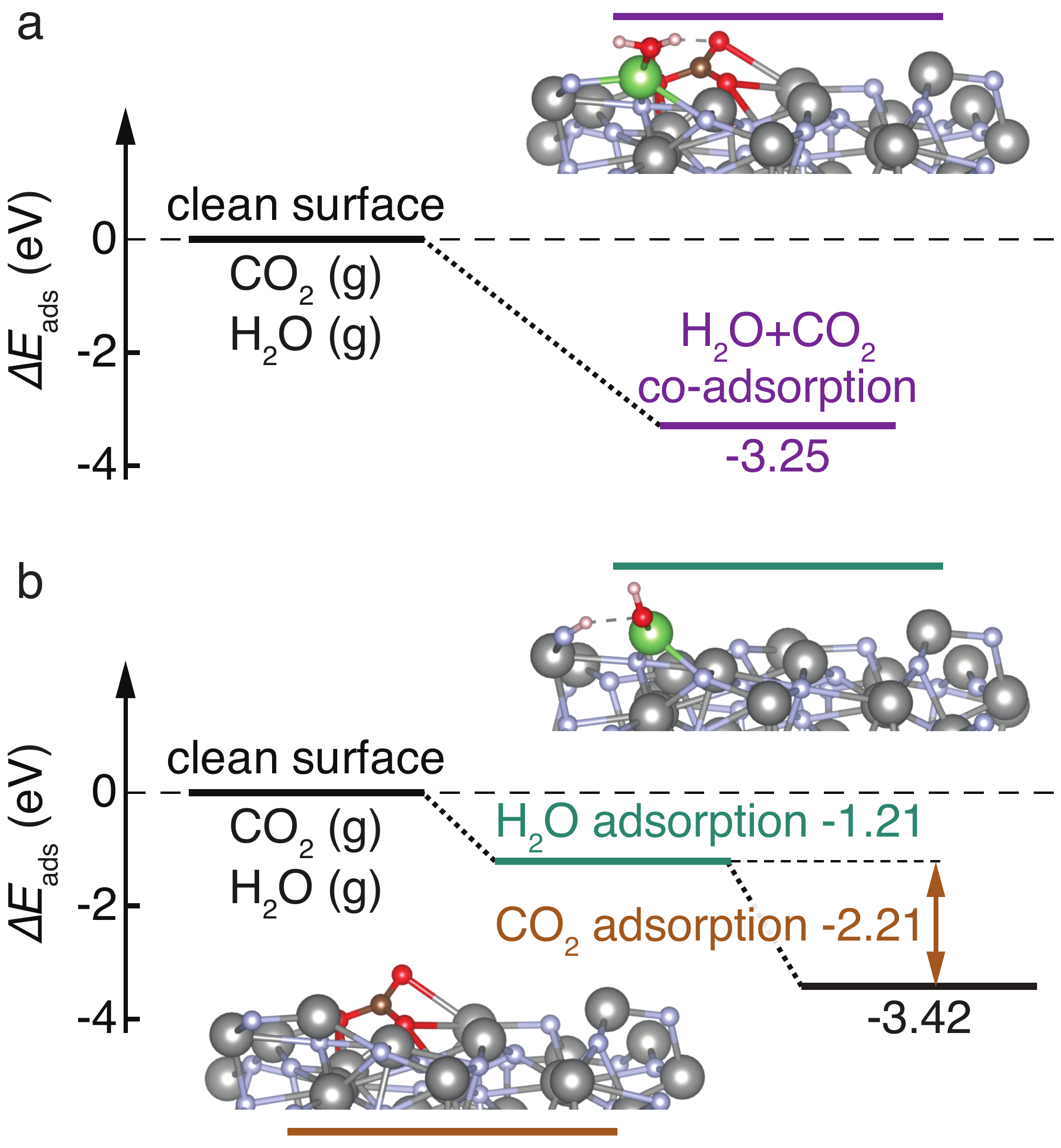}
\caption{$\mathbf{\Delta E_\mathrm{ads}}$ for adsorption of \ch{H2O} and \ch{CO2}: (a) co-adsorption, and (b) infinitely separated adsorption. The co-adsorption shows $\mathbf{\Delta E_\mathrm{ads}}$=--3.25~eV (in purple). The infinitely separated \ch{H2O} adsorption ($\mathbf{\Delta E_\mathrm{ads}}$=--1.21~eV, in green) and \ch{CO2} adsorption ($\mathbf{\Delta E_\mathrm{ads}}$=--2.21~eV, in brown) show the combined adsorption energy of $\mathbf{\Delta E_\mathrm{ads}}$=--3.42~eV, more negative than the co-adsorption. The insets show schematic diagrams for adsorption structures. Atomic species not directly involved in the adsorption are greyed out.}
\label{pathways}
\end{figure}

Having previously demonstrated the pronounced reactivity of \llzo{} surface with \ch{H2O} and \ch{CO2}, we verify  the possibility of  \ch{H2O} and \ch{CO2} reacting concurrently  with the off-stoichiometric Li-terminated (010) cut.  

We investigate the co-adsorption of \ch{H2O} and \ch{CO2} to mimic the hydroxide-mediated carbonation proposed in Ref.~\citenum{sharafiImpactAirExposure2017}, but here mediated explicitely by the \llzo{} surfaces. We adsorbed one \ch{H2O} molecule on the most favorable site Li[3O], and then found the most favorable site to adsorb one \ch{CO2} molecule on the same slab, see inset in Figure~\ref{pathways}(a). In the co-adsorption model, the most favorable \ch{CO2} site is O[LiLaZr] as for the case of individual \ch{CO2} adsorption (see Section~\ref{sec:CO2}).

Next, we take the individual adsorption cases of \ch{H2O} and \ch{CO2} to represent the independent adsorption reactions (Figure \ref{pathways}(b)). This situation corresponds to the infinitely separated adsorption of \ch{H2O} and \ch{CO2} molecules, with the two molecules not interacting with each other.
Figure \ref{pathways}(a) shows $\Delta E_\mathrm{ads}$ of the co-adsorption of of \ch{H2O} and \ch{CO2}, while Figure \ref{pathways}(b) shows the $\Delta E_\mathrm{ads}$s for the infinitely separated  of \ch{H2O} and \ch{CO2} molecules.

The composed $\Delta E_\mathrm{ads}$s  of the individual adsorptions energies of \ch{H2O} ($\Delta E _\mathrm{ads}\sim$--1.21~eV) and \ch{CO2} ($\sim$--2.21~eV) is $\sim$--3.42~eV, which is more negative than the energy of co-adsorption, i.e., $\sim$--3.25~eV.  In the co-adsorption model of Figure \ref{pathways}(a), the adsorbed \ch{H2O} does not easily deprotonate, as opposed to the incipient \ch{H2O} dissociation (at the same site) when adsorbed individually as in Figure \ref{pathways}(b). These evidences clearly inform that individual reactions of \ch{H2O} and \ch{CO2} with \llzo{} compete favorably against the concurrent co-adsorption of \ch{H2O} and \ch{CO2}. Therefore, the surface contamination of \llzo{} will be mostly driven by independent reactions of direct hydration and direct carbonation, as opposed to the hydroxide-mediated carbonation proposed in Ref.~\citenum{sharafiImpactAirExposure2017}.

\section{Discussion}
Using slab calculations in the framework of \emph{ab initio} thermodynamics, we have explored the reactivity  of  the complex oxide \llzo{} with two ubiquitous molecules: \ch{H2O} and \ch{CO2}. It is noted that all the adsorption processes investigated are barrier-free, which indicates fast reaction rates of \ch{H2O} and \ch{CO2} with \llzo{}. A quantitative evaluation of the reaction rates requiring microkinetic models is beyond the scope of this study.

The exposure of \llzo{} to both \ch{H2O} and \ch{CO2} has been documented by preliminary computational and experimental studies.\cite{galvenInstabilityLithiumGarnets2012,larrazCubicPhasesGarnettype2013,sharafiImpactAirExposure2017,sharafiSurfaceChemistryMechanism2017} 

Using  X-ray photo-emission spectroscopy (XPS) depth-profiling \citeauthor{sharafiImpactAirExposure2017} showed the formation of \ch{Li2CO_3} layers as thick as  $\sim$5-10~nm  on the exterior of \llzo{} particles.\cite{sharafiImpactAirExposure2017}   From this evidence, the authors proposed several chemical reactions  (Eq.s~\ref{eq:carbonation2}--\ref{eq:carbonation}) between \llzo{}, \ch{CO2} and \ch{H2O}, whose energetics ($\Delta G$s) were verified by first-principles calculations.\cite{sharafiImpactAirExposure2017}  Nevertheless, the proposed reaction mechanisms of \llzo{} carbonation are still elusive requiring in-depth investigations, which justify this  endeavour.
\begin{align}
\ce{
8Li_7La_3Zr_2O_{12} + CO2(g) & ->[$\Delta G\sim\text{0.17 eV/CO}_2$]  8Li_{7 - 1/4}La_{3}Zr_{2}O_{12-1} + Li2CO3(s)  \label{eq:carbonation2} \\
8Li_7La_3Zr_2O_{12} + H2O (g) & ->[$\Delta G\sim\text{--0.34~eV/H}_2\text{O}$]   8Li_{7 -1/8}H_{1/8}La_{3}Zr_{2}O_{12} + LiOH(s)\label{eq:protonation} \\
8Li_7La_3Zr_2O_{12} + H2O(g) & ->[$\Delta G\sim\text{0.85~eV/H}_2\text{O}$]  8Li_{7-1/4}La_{3}Zr_{2}O_{12-2} + 2LiOH(s) \label{eq:protonation2}\\
LiOH(s) + 1/2CO2(g) & ->[$\Delta G\sim\text{--0.70~eV/CO}_2$]  1/2Li2CO3(s) + H2O(g)  \label{eq:carbonation}
}
\end{align}
The $\Delta G$s in Eq.s~\ref{eq:carbonation2}--\ref{eq:carbonation} are from Ref.~\citenum{sharafiImpactAirExposure2017}.
In Eq.~\ref{eq:carbonation2}, \ch{Li2CO3} is directly formed via carbonation of \llzo{}, but with a predicted positive $\Delta$G.\cite{sharafiImpactAirExposure2017}
Compared to the unfavorable  direct carbonation of \llzo{} (Eq.~\ref{eq:carbonation2}),\cite{sharafiImpactAirExposure2017} the hydroxide-mediated carbonation was interpreted as two favorable reactions: \emph{i}) the protonation of \llzo{} (Eq.~\ref{eq:protonation})  with \ch{H+} exchanging with \ch{Li^+}, followed by \emph{ii}) the conversion of \ch{LiOH} and \ch{CO2} into \ch{Li2CO3} as in Eq.~\ref{eq:carbonation}. Although with unfavorable $\Delta$G ($\sim$0.85~eV/\ch{H2O}), another mechanism proposed was the direct hydration of \llzo{} of Eq.~\ref{eq:protonation2}, which is followed by carbonation (of Eq.~\ref{eq:carbonation}). Therefore, carbonate contamination on \llzo{} was proposed to form via the hydroxide-mediated pathway (Eq.s~\ref{eq:protonation} and \ref{eq:carbonation}) rather than direct carbonation (Eq.~\ref{eq:carbonation2}).\cite{sharafiImpactAirExposure2017}

In Section~\ref{sec:H2O} and Table~\ref{tab:loading} we have amply demonstrated  the favorable adsorption of \ch{H2O} on the \llzo{} surfaces, with the spontaneous \ch{H2O} dissociation on exposed \ch{Li_{Surface}} sites as \ch{Li_{Surface}-O-H}, thus forming \ch{LiOH} species. 
Our calculations clearly suggest that the formation of \ch{LiOH} in \llzo{} is mediated by its reactive surfaces, which ``catalyzes'' the nucleation process of lithium hydroxide moieties. Indeed, XPS of \llzo{} particles have detected signatures of \ch{OH^-} groups, \ch{LiOH} and even at relatively high temperatures (400--500~$^{\circ}$C).\cite{sharafiSurfaceChemistryMechanism2017} 

Another pathway for \ch{LiOH} formation  proceeds through the protonation of \ch{Li-O} moieties exposed at the \llzo{} surfaces, which we also predicted as a spontaneous process. The protonation of other \ch{M-O} (with \ch{M} = \ch{La} or \ch{Zr}) moieties at the \llzo{} surfaces may also form other adventitious species, whose signatures have not been reported yet. However, the availability of free protons at the surface (resulting from readily dissociated \ch{H2O} molecules) can then be exchanged with \ch{Li^+} in \llzo{} as previously reported by thermo-gravimetric analysis mass-spectroscopy, neutron investigations, and as focused ion beam secondary ion mass spectrometry.\cite{larrazCubicPhasesGarnettype2013,galvenInstabilityLithiumGarnets2012, bruggeGarnetElectrolytesSolid2018} Alternatively, proton species formed on the  \llzo{} surfaces may intercalate directly into vacant \ch{Li^+} sites of the bulk structure. Note, the high \ch{Li}-ion transport \llzo{} facilitates \ch{Li^+}/\ch{H^+} exchange.

\begin{figure*}[!ht]
\centering
\includegraphics[width=1.0\textwidth,clip]{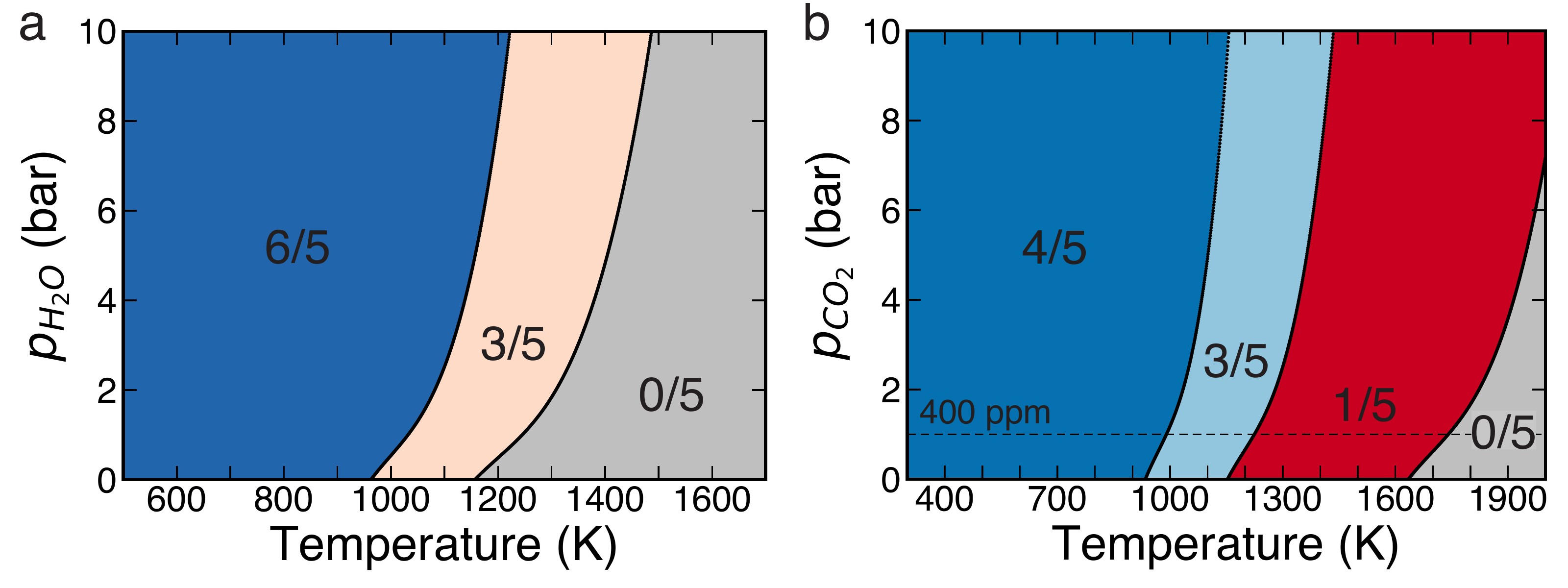}
\caption{Surface phase diagrams of \llzo{} at different temperatures and partial pressures of (a) \ch{H2O} and (b) \ch{CO2}, respectively. Fractions of  monolayer (ML) and solid color denote the most favorable coverage scenarios within each range of temperature (in K) and partial pressure (in bar). The dashed line marks \ch{CO2} level of 400 ppm at partial pressure of 1 bar. This line is used to indicate the thermodynamic favorability of adsorbing \ch{CO2} from dilute streams, which is of relevance in \ch{CO2} utilization.}
\label{surface_diagram}
\end{figure*}

\ch{CO_3^{2-}} contamination of the \llzo{} particle exteriors has been previously documented through Raman and XPS measurements.\cite{sharafiImpactAirExposure2017,sharafiSurfaceChemistryMechanism2017}
Our calculations demonstrated spontaneous \ch{CO2} adsorption and carbonate formation with the exposed oxygen atoms, i.e., \ch{O_{Surface}-Li_{Surface}} or \ch{O_{Surface}-La_{Surface}} (see Section \ref{sec:CO2} and Table~\ref{tab:loadingCO2}) at the \llzo{} surfaces. This is in striking contrast to the unfavorable reaction of direct carbonation proposed in Eq.~\ref{eq:carbonation2}.\cite{sharafiSurfaceChemistryMechanism2017} Importantly, we show that the formation of \ch{Li2CO3} does not require the availability of \ch{LiOH} with \ch{CO2}, but it is directly catalyzed by the \llzo{} surfaces. Our surface calculations show that carbonates are not limited to \ch{Li2CO3} species, instead incipient \ch{CO3^{2-}} species may be bonded to other metal ions exposed at the surfaces of \llzo{}, including \ch{Li} and \ch{La}. Therefore, direct carbonation of \llzo{} is thermodynamically favorable and barrier-free.
Notably, the co-adsorption of \ch{H2O} and \ch{CO2} appear less favorable than their individual adsorptions.

We have demonstrated the thermodynamically favorable processes of \ch{H2O} and \ch{CO2} decomposition on the \llzo{} particles.  We use this knowledge to prescribe external conditions (temperature and pressure) which curb the \ch{LiOH} and \ch{Li_2CO_3}  contamination of \llzo{} during its synthesis. Figure~\ref{surface_diagram} shows the computed surface phase diagrams of \llzo{} as functions of temperature and partial pressures of \ch{H2O} and \ch{CO2}. The derivation of the surface phase diagrams requires the minimization of the grand-potential energy of Eq.~\ref{eq:2_g} as the surface coverage, temperature, and pressures are varied.\cite{canepaParticleMorphologyLithium2018}

From Figure~\ref{surface_diagram}, one clearly sees that scenarios of high-\ch{H2O} or \ch{CO2} coverages are achieved at relatively high partial-pressures ($\sim$2--10~bars) and low temperatures ($<$1,000~K). Our predictions suggest that at atmospheric pressure ($\sim$1~bar) and at temperature above $\sim$1,260~K, \llzo{} particles will be free of water, protons and \ch{LiOH} species. More pernicious appears the carbonation of \llzo{}, which clearly requires higher temperatures above $\sim$1,730~K for its complete elimination at atmospheric pressures (see Figure~\ref{surface_diagram}(b)). These findings are representative of the real \llzo{} powders or the surfaces of sheet (\llzo{} composites), as the surface investigated dominates the \llzo{} Wulff shapes.\cite{awakaSynthesisStructureAnalysis2009,canepaParticleMorphologyLithium2018}

The synthesis of \llzo{} is commonly carried out via the solid-state method, by calcination in ``dry'' air at temperatures $\sim$1,000~$^{\circ}$C ($\sim$1,273~K).\cite{sharafiControllingCorrelatingEffect2017,sharafiImpactAirExposure2017,sharafiSurfaceChemistryMechanism2017}
Procedures of densification of \llzo{}  via hot-press (or spark-plasma) are  regularly utilized to improve contact between particles and increase the electrolyte ionic conductivity. 
The surface phase diagrams of Figure~\ref{surface_diagram}, clearly show that both synthesis and sintering of \llzo{} should avoid air-exposure.
Notably, Figure~\ref{surface_diagram} informs that post-synthesis heat treatments (above 1,000$^{\circ}$C) are beneficial to desorb \ch{H2O} and eliminate other contaminants. Indeed, typical sintering temperatures for \llzo{} are in the range of  1,000--1,300$^{\circ}$C ($\sim$1,300--1,600~K).\cite{sharafiImpactAirExposure2017,sharafiControllingCorrelatingEffect2017}  Higher temperatures than 1,300$^{\circ}$C may lead to significant Li (and proton) loss.\cite{sharafiControllingCorrelatingEffect2017} However, such elevated temperatures may still prevent full removal of \ch{Li2CO3} at the surfaces. {\ch{Li}} loss during the {\llzo{}} preparation is typically mitigated by adding addition {\ch{LiOH}} and {\ch{Li2CO3}}. Furthermore, we speculate that the addition of  {\ch{LiOH}} and {\ch{Li2CO3}} will not be disruptive to the preparation of {\llzo{}}, and these two compounds may also limit their formation when {\llzo{}} is exposed to humid air.

Although the effect of partial pressure of \ch{H2O} and \ch{CO2} appears less significant than temperature with respect \llzo{} contamination, synthesis under ultrahigh vacuum conditions ($\sim$1$\times$10$^{-6}$~bar) and at $\sim$800~K guarantee water-free \llzo{} surfaces  and sparsely contaminated by \ch{Li2CO3} species.  Clearly, these conditions appear  unpractical, but glove-box synthesis at pressures within mbars ranges (see Figures~\ref{multi_H2O} and \ref{multiple_CO2}) may be more accessible and deserve further investigations.

Figure~\ref{surface_diagram} indicates external conditions which favor (010) \llzo{} surfaces decorated by dissociated \ch{H2O} and carbonate species. While these surfaces may not be desired for energy storage applications, the strong adsorption of \ch{CO2} and \ch{H2O} suggests that the off-stoichiometric (010) surface of \llzo{} can serve as a support in catalysis. Using two examples of water gas shift and \ch{CO2} utilisation, we elaborate further below. Water-gas shift is typically catalyzed by noble metal nanoparticles stabilized by oxide supports. Experimental and theoretical studies have indicated that the oxide supports facilitate \ch{H2O} dissociation, \cite{zhaoImportanceMetaloxideInterfaces2017,nelsonHeterolyticHydrogenActivation2020} a critical elementary step in this reaction. The favorable adsorption of \ch{H2O} and its spontaneous dissociation thus makes \llzo{} a promising support for the water-gas shift reaction. Indeed, there is a concerted effort to design catalytic systems which operate under dilute streams of \ch{CO2}. \cite{senHydroxideBasedIntegrated2020,omodolorDualFunctionMaterialsCO2020} Figure~\ref{surface_diagram} illustrates that at ~ 500-600K, 4/5 ML of \ch{CO2} is adsorbed on \llzo{} even at \ch{CO2} pressures as low as 400 ppm. Hence, the strong adsorption strength of \ch{CO2} on \llzo{} makes this material a promising support for dual-functional catalysts, which adsorb \ch{CO2} from dilute streams and further hydrogenate it to chemicals, such as methanol and ethylene over metal nanoparticles dispersed on \llzo{}.

\section{Conclusions}

In summary, using a robust \emph{ab initio} thermodynamic framework, we investigated the complex surface reactivity of  \llzo{} toward ubiquitous \ch{H2O} and \ch{CO2}.  Li-terminated off-stochiometric surfaces of \llzo{} readily react with \ch{H2O} and \ch{CO_2}, which promotes the direct formation of \ch{LiOH} and \ch{Li2CO3}. We demonstrated that synthesis within mbar pressure can lead to water-free \llzo{} surfaces and reduce drastically the amount of \ch{Li2CO3} formed.  
Our in-depth analysis provides strategies and opportunities to improve the synthesis conditions of \llzo{} and similar complex oxides of high technological relevance in batteries and for catalysis. The strong adsorption of \ch{CO2} under conditions of low partial pressure and of \ch{H2O} indicates the potential of \llzo{} as a suitable catalyst support for the hydrogenation of dilute \ch{CO2} streams and the water-gas shift reaction respectively.
These finding are general and provides thumb rules to analyze the reactivity of complex oxides toward two common molecules, i.e., \ch{H2O} and \ch{CO2}.

\begin{acknowledgement} Y.L.\ and P.C.  are grateful to the ANR-NRF NRF2019-NRF-ANR073 Na-MASTER. P.C.\ acknowledges funding
from the National Research Foundation under his NRF Fellowship
NRFF12-2020-0012. T.\ S.\ C. gratefully acknowledges funding from the Ministry of Education Academic Research Fund Tier-1 RS04/19 and RG5/21. A.\ M.\ P.\ acknowledges Nanyang Technological University for a Research Scholarship. The
computational work was performed on resources of the National
Supercomputing Centre, Singapore (\url{https://www.nscc.sg}). 
\end{acknowledgement}

\begin{suppinfo}
The supporting information includes: \emph{i}) Benchmark tests on the effect of  vdW corrections, \emph{ii}) convergence tests of slab calculations, \emph{iii}) structure of the slab unit cell, \emph{iv}) tests on single-point energy cutoffs, \emph{v}) the description of the metal sites' local environment and \ch{H2O} dissociation, \emph{vi}) calculation results for forced \ch{H2O} dissociation at high coverage, \emph{vii}) geometry of surface carbonates formed at the surfaces and \ch{Li2CO3}, and \emph{viii})  Mulliken charges of sites post-adsorption. 
\end{suppinfo}

\bibliography{biblio}

\end{document}